\documentclass[prb,twocolumn,amsmath,amssymb]{revtex4}
\usepackage{amsmath}
\usepackage{graphicx}
\usepackage{rotating}
\usepackage{units}
\usepackage{color}
\usepackage{ulem}

\begin{document}

\title{Nanoscale magnetometry using a single spin system in diamond}

\author{R. S. Said$^{1}$, D. W. Berry$^{2}$, J. Twamley$^{1}$}
\affiliation{$^1$Centre for Engineered Quantum Systems, Department of Physics \& Astronomy, Faculty of Science, Macquarie University, Sydney, NSW 2109, Australia.\\
$^2$Institute for Quantum Computing and Department of Physics \& Astronomy, University of Waterloo, Ontario \mbox{N2L 3G1}, Canada.}

\begin{abstract}
We propose a protocol to estimate 
magnetic fields using a single nitrogen-vacancy (N-V) center in diamond, where the estimate precision scales inversely with time, $\delta B\sim 1/T$, rather than the square-root of time, $\delta B\sim 1/\sqrt{T}$.
The method is based on converting the task of magnetometry into  phase estimation, performing quantum phase estimation on a single N-V nuclear spin using either adaptive or nonadaptive feedback control, and the recently demonstrated capability to perform single-shot readout within the N-V [P. Neumann \textit{et al.}, Science {\bf 329}, 542 (2010)].
We present numerical simulations  to show that our method provides an estimate whose precision scales close to $\sim 1/T$ ($T\sim$ the total estimation time), and moreover will give an unambiguous estimate of the static magnetic field experienced by the N-V. By combining this protocol with recent proposals for scanning magnetometry using an N-V, our protocol will provide a significant decrease in signal acquisition time while providing an unambiguous spatial map of the magnetic field.
\end{abstract}

\maketitle

\section{Introduction}

A highly sensitive magnetic field sensor, that can operate at room temperature and has atomic spatial resolution may revolutionize many nanotechnologies, for instance in medical and biological technologies, advanced material sciences, spintronics, and quantum computing. Toward the practical realization of such a device, nanoscale magnetometry experiments in solids have been realized using single nitrogen-vacancy (N-V) centers in diamond.\cite{Maze2008,Balasubramanian2008} These experiments detect very weak magnetic fields, $\sim$ 3 nT at kilohertz frequencies, and can locate a nearby electronic spin with a spatial resolution of $\sim$ 5 nm by utilizing electron spin dynamics of a defect center in a diamond nano-crystal, as illustrated in Fig. \ref{nvtip}. This defect center possesses remarkable properties for magnetic sensing: It can be individually addressed and optically polarized and measured, and it maintains spin coherence at room temperature for considerable periods of time.\cite{Wrachtrup2006,Wrachtrup2006b} However, the current magnetometry precision is limited by standard statistical fluctuations, namely the shot-noise limit.\cite{Taylor2008} In this limit the precision of the magnetic field estimate scales as $\delta B\sim 1/\sqrt{T}$, where $T$ is the total time needed to acquire the estimate. 

This scaling is because $T/\tau$ independent measurements are made over a short time $\tau$. This yields an uncertainty in the magnetic field of\cite{Taylor2008} $\delta B\approx (\hbar/g\mu_B)(1/\sqrt{\tau T})$. In principle, if one were to use a measurement over the entire time interval $T$, then one would have a measurement with uncertainty scaling as $\delta B \sim 1/T$. This is the best precision possible for a measurement over this time interval, and is equivalent to the Heisenberg limit for phase measurement.\cite{Berry2009} There are two problems preventing measurements with precision scaling as $1/T$. The first is spin-spin relaxation; performing a single measurement beyond the dephasing time $T_2$ does not yield an increase in precision. The second is that performing a measurement over a longer time may result in ambiguities. That is, the magnetic field causes the spin to rotate more than once, and the number of rotations cannot be determined from the measurement. Experimental advances have increased $T_2$, and there are proposals to extend $T_2$ to the order of a second.\cite{Hall2010}

In this work, we address the second problem, and present a method to achieve Heisenberg-like scaling of the precision for measurement times smaller than $T_2$ while eliminating any ambiguities in the estimation. This will allow faster acquisition of a magnetic field map for a preset precision. In summary, we adapt  a more generalized quantum phase estimation algorithm (gQPEA),\cite{Higgins2007} to instead estimate the phase generated by an unknown $Z-$rotation of the Bloch sphere of a qubit (atomic two level system), via Ramsey interferometry. The gQPEA phase estimation algorithm was initially developed using the framework of optical interferometry to estimate an unknown phase acquired when a photon passes through a static phase shifter, and it has been experimentally demonstrated using linear optical methods.\cite{Higgins2007,Higgins2009} The protocol can make use of either adaptive\cite{Berry2006,Higgins2007} or nonadaptive\cite{Higgins2009,Berry2009} controls to yield unambiguous estimates of the phase with a precision which scales inversely with the overall measurement time, i.e.,  Heisenberg-like scaling. Our protocol makes use of single-shot measurements (SSM) of the spin of the atomic system, and we generalize to the case when the visibility of such measurements may be significantly below 100\%.

We numerically simulate our magnetometry protocols under both ideal and realistic measurement conditions, taking into account atomic decay and dephasing. In the ideal case, when we have perfect SSM visibility, we predict that the precision of the magnetic field estimate {has better scaling than the} shot-noise limit; i.e., the precision scales as $\delta B \sim 1/T^\beta,\;\;1/2< \beta <1$. We describe this type of scaling as {\it sub-shot-noise scaling}. Furthermore, we analyze the performance of these protocols when the SSM visibility can be quite low, and surprisingly find that sub-shot-noise scaling in the estimate precision is still possible.

We begin by reviewing Ramsey interferometry in a two-level system to estimate an unknown phase rotation. Section III  adapts the gQPEA to atomic systems to operate within the  Ramsey interferometry cycle.  Section IV presents the results of numerical simulations for the adaptive and nonadaptive protocols with varying single-shot measurement visibilities. 

We surprisingly discover that, for reduced SSM visibility, the nonadaptive method performs better than the adaptive method. Furthermore, we find that the nonadaptive method can achieve a scaling close to $\delta B\sim 1/T$ even when the SMM visibility is substantially reduced.  Hence, we focus our attention on optimizing the nonadaptive protocol to work as efficiently as possible in  experimentally realistic conditions where one may not have perfect single-shot measurement capability. We finally discuss our conclusions and possible future directions.

\section{Spin interferometry}

Before reviewing Ramsey spin interferometry, we clarify some important definitions used throughout our paper. The resource required  to obtain an estimate of the magnetic field to a preset precision is essentially the total interaction time between the atomic probe and the unknown magnetic field. We denote this total interaction time by $T$. {For shot-noise scaling the uncertainty of the estimate scales as $\sim 1/\sqrt{T}$, sub-shot-noise scaling is $\sim 1/T^\beta$ for $1/2 < \beta < 1$, and Heisenberg-like scaling is $1/T$.}

We now consider a two-level spin system which experiences a static unknown magnetic field $B_z$. Through a pulse sequence, which is analogous to Ramsey atomic interferometry, we can convert the problem of estimating this unknown magnetic field strength into the problem of estimating the unknown phase $\phi\propto\lambda_g B_z t$ acquired by the spin during a time $t$, where $\lambda_g$ is the gyromagnetic ratio of the spin, and $t$ is predetermined accruing time between the spin and the magnetic field. Our protocol will estimate the value of the accrued phase $\phi$. 

\begin{figure}[tp] 
\begin{center}
\setlength{\unitlength}{0.9cm}
\begin{picture}(8.5,8.5)
\put(.45,0){\includegraphics[height=8.5\unitlength]{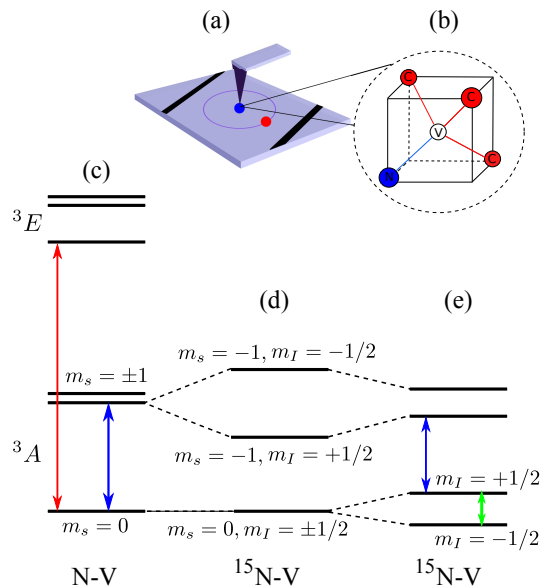}}
\end{picture}
\end{center}
\vspace{.5cm}
\caption{(Color online) (a) A single spin of an N-V center (blue point), attached to the tip of an atomic force microscope probe, detects a weak magnetic field from another spin (red point). The `blue'  probe's dynamics {are} controlled by microwave wires (two black stripes). (b) Structure of the center, consisting of a substitutional nitrogen impurity N (blue point) adjacent to a carbon vacancy V (white point), tetrahedrally coordinated to each other and three nearby carbon atoms C (red points).\cite{Lenef1996} The impurity can be a $^{14}$N($I=1$) or $^{15}$N($I=1/2$) atom. (c) Electron spin levels of the center. The ground state $^3A$ is a triplet ($S=1$) with a $2.88$ GHz zero-field splitting between $m_s=0$ and $m_s=\pm1$. Standard Ramsey interferometry performed in the electron spin is the basis for the operation of many magnetometry protocols.\cite{Maze2008,Balasubramanian2008,Taylor2008} (d) Hyperfine level splitting due to a coupling to the $^{15}$N nuclear spin \cite{Rabeau2006} in the absence of nuclear Zeeman effect. (e) Degeneracy ($m_s=0,m_I=\pm 1/2$) can be lifted to allow optically detected nuclear spin resonance. Red, blue, and green arrows represent optical, microwave, and radio-frequency transitions, respectively. The qubit proposed to probe the weak magnetic field is the `green' transition.}
\label{nvtip}
\end{figure}

We will choose the two interferometric basis states $\{|0\rangle,|1\rangle\}$, whose superpositions will precess in the magnetic field, to be  the two nuclear hyperfine states $|m_s=0, m_I=\pm 1/2\rangle$, as shown by the green transition in Fig. \ref{nvtip}(e). We choose these for two reasons: First, as recently demonstrated,\cite{Neumann2010b} one can perform single-shot measurements of the nuclear spin state associated with these basis states, and second, these nuclear spin states possess very long coherence times.  The hyperfine splitting between the two states  $|m_s=0, m_I=\pm 1/2\rangle$, arises from the coupling between the electron spin of the $^{15}$N-V center ($S=1$) in its ground state, and the $^{15}$N nuclear spin ($I=1/2$). However, another choice might be to choose the shorter lived electronic spin states $\{ |m_s=0\rangle,|m_s=\pm 1\rangle \}$ of the N-V center,\cite{Steiner2010} as depicted by the blue transitions in Fig. \ref{nvtip}(c).

{Our protocol requires one to perform many single-shot Ramsey measurements using a range of time durations}. We summarize each sequence as follows: The spin is initialized to a state $|0\rangle \equiv |m_s=0,m_I=-1/2 \rangle$, which can be pictured as a vector along the $+Z$ direction in the Bloch sphere. This initialization can be carried out in the N-V system by optical cycling, near resonance on the optical transition ${}^3A-{}^3E$. We then apply a $\pi/2$-pulse on resonance with the $|0\rangle-|1\rangle$ transition, and this produces a pure superposition state, written in density matrix form as 
\begin{eqnarray}
\rho_0 = \frac{1}{2}\left( |0\rangle\langle 0| + i |0\rangle\langle 1|  - i |1\rangle\langle 0| + |1\rangle\langle 1|\right). \label{initialdensitymatrix}
\end{eqnarray}
That is, the initial vector is rotated to the equator of the Bloch sphere, and that rotation is assumed to be implemented in a very short time. This rotation can be realized by applying a radio-frequency field if we use the nuclear spin to encode the interferometric basis states, or by using a microwave field in the case of electronic spin states. The spin system then undergoes free precession due to the effect of the unknown external static magnetic field $B_z$. For some predetermined period of time $t$, the spin system accrues a relative dynamical phase $\phi$, proportional to the magnetic field strength, and rotates the Bloch vector around the $Z$ axis in the $X-Y$ plane.

During this precession we have the Hamiltonian ${\cal H}_f = \lambda_g B_z \hat{\sigma}_z $, and including  decay and dephasing, we can describe the evolution via a master equation in the Lindblad form \cite{Havel2003}, as
\begin{eqnarray}
\dot{\rho}(t) &=&  - \frac{i}{\hbar} \left[ {\cal H}_f, \rho(t) \right] + \frac{1}{T_1} {\cal L} (\hat{\sigma}_-, \rho(t)) \cr
&&+ \left( \frac{1}{2 T_2} -  \frac{1}{4 T_1} \right) {\cal L} \left(\hat{ \sigma}_z, \rho(t) \right), \label{Master}
\end{eqnarray}
where ${\cal L}(\hat A,\hat B)=\hat A \hat B \hat A^\dag - \frac{1}{2} \left\{\hat A^\dag \hat A,\hat B \right\}$, $\hat{\sigma}_z = |0\rangle\langle 0| - |1\rangle\langle 1|$, and $\hat{\sigma}_{-}=|0\rangle\langle 1|$. We denote the decay time as $T_1$. Setting $\rho(t=0)=\rho_0$, and $\hbar=1$ for simplicity, we obtain a solution to (\ref{Master}) as
\begin{equation}
\rho(t) = 
\left[
\begin{array}{ccc}
1- \frac{1}{2} \exp({-\frac{1}{T_1}t})  & \frac{1}{2} i \exp({\Lambda^* t}) \\
- \frac{1}{2} i \exp({\Lambda t}) & \frac{1}{2} \exp({-\frac{1}{T_1}t}) 
\end{array}
\right],
\end{equation}
where $\Lambda = 2i\lambda_g B_z - 1/T_2$. We then apply another $\pi/2$-pulse ($\pi/2$ rotation around the $X$ axis), mapping the phase information into population information.  We then perform a single-shot measurement of the population to get a single detection result.  For clarity we call such a single-shot measurement a `click', in contrast to the overall measurement of magnetic field, which includes many of these individual `clicks'. The probability of measuring the system in the state $|0\rangle$ or $|1\rangle$ (+1 or -1 click along the $Z$ axis), conditioned on the unknown field $B_z$, is
\begin{eqnarray}
P(\pm 1 | B_z)  = \frac{1}{2}\left[ 1 \pm    e^ {-t/T_2}  \cos \left( 2\lambda_g B_z t \right)\right].
\label{Conditional_1}
\end{eqnarray}
For a small predetermined accruing time $t=\tau$, where $\tau \ll T_2$, and repeating the standard prepare-evolve-measure procedure $M$ times, the total amount of time resource expended is $T=M \tau$. This naive procedure provides a standard deviation of the estimate that scales as\cite{Taylor2008} $\delta B_z \sim 1/\sqrt{T}$. {This is the shot-noise limit for measurement precision. In what follows, we will show that by adapting the gQPEA protocol,\cite{Higgins2007,Higgins2009} which involves acquisition of `clicks' over varying time durations, we can obtain an unambiguous magnetic field estimate with precision that scales better than the shot-noise limit.}

\section{Estimation method}
To {provide} clear insight into the estimation method, we recast (\ref{Conditional_1}) into
\begin{eqnarray}
P(\pm 1 | \phi)  = \frac{1}{2} \left[ 1 \pm e^ {-t/T_2} \cos \left( \phi \right)\right],
\end{eqnarray}
where the system phase $\phi=2 \lambda_g B_z t$, is the unknown and constant real quantity we would like to estimate. The obtained estimate itself, after the estimation procedures, is denoted by $\hat{\phi}$ to differentiate it from the actual underlying system phase.
Following {Ref.\ \cite{Berry2009}}, we expand the probability distribution for the system phase after the $m$-th single-shot measurement as
\begin{eqnarray}
P_m(\phi)=\sum_{j} b_{j,m} e^{ i\, j \phi}\;,
\end{eqnarray}
and can compute the expectation value for $\exp (i \phi)$ after the $m$-th click as $\langle e^{i \phi} \rangle_m = \int P_m(\phi) e^{i\phi} d\phi$. 

{If there is no initial knowledge about the phase, the initial probability distribution $P_{m=0}(\phi)$ is flat}. After the first detection, $m=1$, we have one bit of information about the system phase that can be used to calculate the three nonzero coefficients $b_{j=\{0,\pm 1\},m=1}$. After $m$ detections are made, a vector of detection results $\vec{u}_m=\{u_{1},u_{2},\ldots,u_{m}\}$ is obtained, where each $u_{i}(i=1,\ldots, m)=\pm 1$. {Using Bayes' rule, the conditional probability distribution for the system phase given the next detection result is}
\begin{eqnarray}
 P \left(\phi | \vec{u}_{m+1}\right)  \propto P \left(\pm 1 | \phi\right) P\left(\phi | \vec{u}_{m}\right),
\end{eqnarray}
{with $P(\phi|\vec{u}_m)\equiv P_m(\phi)$.}
It gives an update formula for the unnormalized coefficients of the probability distributions $P_{m}(\phi) \mapsto P_{m+1}(\phi)$ as
\begin{eqnarray}
\tilde b_{j,m+1}= b_{j,m} + u_{m+1} e^{-\frac{t}{T_2}}  \left( \frac{b_{j-1,m} + b_{j+1,m}}{2} \right),
\end{eqnarray}
where the normalized coefficient is $b_{j,m+1}=\tilde b_{j,m+1} / 2\pi \tilde b_{0,m}$.

The estimate of the system phase $\hat{\phi}$ after a total of $m=M$ detections is then obtained by taking the argument of the expectation value of $e^{i\phi}$,
\begin{align}
\hat{\phi} &= \arg \langle e^{i\phi} \rangle_{M} = \arg \left( \int P\left( \phi | \vec{u}_{m=M} \right) e^{i\phi} d\phi \right) \nonumber \\ &= 
\arg \left( b_{-1,M} \right).
\end{align}

Averaging over a number of trials (or samples) $\mathbf{S}$, we obtain { an estimate of the Holevo variance $V_H$ \cite{Holevo1984}, corresponding to the square of the uncertainty in the phase estimate,} as 
\begin{eqnarray}
V_H=\left( \overline{|2\pi b_{-1,M}|} \right)^{-2} -1.
\end{eqnarray}
The case when $\delta B \sim 1/\sqrt{T}$, (usual shot-noise limit) can be recast as $V_H T\sim C$ {(a constant),} while the limit where $\delta B\sim 1/T$ can be recast as $V_H T\sim 1/T$. Below we will identify sub-shot-noise scaling where $V_H T\sim 1/T^\beta,\;\; 1> \beta>0$.

Rather than using standard Ramsey interferometry for phase estimation we make use of another protocol, originally devised in an optical setting, which uses multiple accumulations of the unknown phase to achieve near Heisenberg scaling of the {uncertainty in the phase estimate}.\cite{Higgins2007,Higgins2009,Berry2009} The multiple accumulations correspond to a prolonged accruing time, which provides more information about the phase.

Applying this multi-accumulation concept to our problem, we modify the expression of (4) by replacing $t=\tau$ with $t=\tau_k\equiv 2^k\tau$:
\begin{eqnarray}
P(\pm 1 | \phi)  = \frac{1}{2}\left[1  \pm e^ {-\tau_k/T_2} \cos \left( 2^k\phi \right)\right],
\end{eqnarray}
where now $\phi=2\lambda_g B_z \tau$. {Prolonging the accruing time increases the rotations of the state vector in the $X-Y$ plane} and consequently introduces a modulo $2\pi$ ambiguity into the estimate. We eliminate this ambiguity through adjustment of the parameter $k$, and the application of an additional  effective phase shift $\Phi$ to the system. This can be achieved via the active application of an external magnetic field or passively, through simply performing the second Ramsey $\pi/2$ rotation about a rotated axis on the $X-Y$ plane of the Bloch sphere instead of the $X-$axis. This additional phase shift can be chosen adaptively (as feedback) or nonadaptively (as a predetermined phase increment), such that
\begin{eqnarray}
P(\pm 1 | \phi,\Phi)  = \frac{1}{2}\left[ 1 \pm  V(\tau_k) \cos \left( 2^k\phi - \Phi \right)\right],
\label{DetectionProbFeedback}
\end{eqnarray}
{where $V(\tau_k) = \exp \left( -2^k \tau/T_2 \right)$ is} the estimation visibility. Given a conditional detection probability we can determine the Fisher information,\cite{OLoan2010}
\begin{eqnarray}
{\cal F}_\phi &=& \sum_{\xi =+1,-1}  \frac{1}{P\left( \xi | \phi,\Phi \right)} \left( \frac{\partial P\left( \xi | \phi,\Phi \right)}{\partial \phi}\right)^2 \nonumber \\
&=& \frac{4^k \sin^2\left(2^k\phi - \Phi\right)}{\exp\left(2^{1+k}\tau/T_2\right) - \cos^2 \left(2^k\phi - \Phi\right)},
\label{Fisher}
\end{eqnarray}
where $\xi$ is a {click} measurement result.

The Fisher information ${\cal F}_\phi$ essentially represents the amount of information about $\phi$ contained in the measurement results $\xi$. 
Maximal information about the system phase can be extracted when we choose values of $k$ that maximize (\ref{Fisher}), for a fixed ratio of $\tau/T_2$.  
Further investigation of (\ref{Fisher}) allows one to show that no useful information regarding the system phase can be inferred when $T=2^k\tau \gg T_2$. This is numerically confirmed by our simulations as presented below. The overall measurement time can be greater than $T_2$, but the uncertainty scales as $1/\sqrt{T}$ for $T>T_2$. Beyond $T_2$ we can only hope to achieve an uncertainty which scales, at best, {like} the shot-noise limit.

In the following, we first describe a protocol for sub-shot-noise magnetometry which uses adaptive feedback, where the value of the control phase $\Phi$ in (\ref{DetectionProbFeedback}) is determined based on previous detection results. Following this we describe a method which is perhaps more suitable for experiments, where the control phase $\Phi$ is regularly incremented without any dependence on the previous detection results. 

\subsection{Adaptive scheme}

For the adaptive control scheme, the control phase $\Phi$ in (\ref{DetectionProbFeedback}) is updated after each detection based on the previous detection results. Ideally, this control phase is altered in such a way so as to minimize the variance of the final phase estimate. However, there is no known method to determine such a phase analytically, and numerical methods are computationally intensive. Instead, we use an adaptive update scheme\cite{BerryThesis} to minimize the variance of the phase estimate after the next detection, based on the information obtained from the previous detections. 

In order to minimize the variance after the next detection, one must maximize the quantity $\cal M$ \cite{BerryThesis},
\begin{eqnarray}
{\cal M} (\Phi_{m})= \frac{1}{2\pi} \sum_{u_{m}=0,1} \left|  \int e^{i\phi} P ( \vec{u}_{m} | \phi ) d\phi\right|.
\end{eqnarray}
The adaptive control phase that maximizes $\cal M$
is one of \cite{BerryThesis}
\begin{eqnarray}
\Phi_0 &=& \arg \left(ba^* - c^*a\right), \cr
\Phi_\pm &=& \arg \left(\sqrt{\frac{c_2\pm \sqrt{c_2^2+|c_1|^2} }{c_1}}\right),
\end{eqnarray}
where 
\begin{eqnarray}
c_1&=&(a^*c)^2-(ab^*)^2+4(|b|^2-|c|^2)b^*c, \nonumber \\
c_2 &=&-2 i {\mathbf Im} (a^2b^*c^*),
\end{eqnarray}
and $a, b$, and $c$ are functions of the $b_{j,m-1}$ expansion coefficients for the probability distribution, 
$a = b_{-1,m-1}$,
$b=b_{-1-2^k,m-1} V(\tau_k)$,
$c= b_{-1+2^k,m-1} V(\tau_k)$. 
Finally, we arrive at the update formula for $b_{j,m}\mapsto b_{j,m+1}$, based on the acquisition of the $(m+1)$-th click measurement $u_{m+1}$, and using (\ref{DetectionProbFeedback}) and the control phase $\Phi_{0,\pm}$ that maximizes $\cal M$, to obtain
\begin{eqnarray}
\tilde b_{j,m+1} &=& b_{j,m} + u_{m+1} V(\tau_k) \nonumber \\
&& \times \left(  \frac{b_{j-2^k,m} e^{-i \Phi_m} + b_{j+2^k,m} e^{i \Phi_m}}{2} \right).
\label{updatingformulafortheprobabilityamplitude}
\end{eqnarray}

For clarity, we enumerate the individual steps in the estimation procedure using the adaptive control phase as follows:

\begin{enumerate}
\item
All parameters are set: the fundamental accruing time $\tau$, the maximum number of multi-accumulations {$2^K$}, the maximum number of detections $M$, the number of trials $\mathbf{S}$, and the dephasing time $T_2$.
\item
For the $\mathbf{s}$-th trial, the protocol starts using the initial $k=K$ and the initial additional phase $\Phi$.
\item
The spin state is initialized to $|0\rangle$.
\item
The $\pi/2$-rotation around the $X$-axis is applied by the external pulse (on-resonance with the $|0\rangle - |1\rangle$ transition) to create the coherent superposition state.
\item
The Bloch vector undergoes free evolution in the $X-Y$ plane for $\tau_k=2^k\tau$, due to the non-varying unknown magnetic field and the additional phase $\Phi$.
\item
Another $\pi/2$-rotation around the $X$-axis is applied to transform the phase information into the spin population. Instead of the active application of the phase shift $\Phi$ in step 5, one can simply rotate the axis of this second Ramsey rotation.
\item
A single-shot measurement is performed to provide the result $\xi=\pm 1$.
\item
The phase estimate $\hat{\phi}$ is inferred and the additional phase $\Phi$ is updated based on the measurement result $\vec{u}_m$.
\item
Steps 3--8 are repeated $M$ times. 
\item
Steps 2--9 are repeated for $k=K, K-1, K-2,\ldots, 0$. 
\item
Steps 2--10 are repeated $\mathbf{S}$ times. In the end, one has $\mathbf{S}$ number of final estimates required to obtain a numerical estimate of the Holevo variance $V_H$.  
\end{enumerate}

To complete a single trial (steps 1-9), the protocol requires a total time resource $T_{af}=M \times (2^{K+1}-1) \tau$. Below we will use $T_2/\tau=10^3$, and taking\cite{Balasubramanian2008} $T_2\sim 2$ ms, we have a fundamental accruing time of $\tau=2\,\mu{\rm s}$.  
For simplicity, we will {from} now on make reference to dimensionless time quantities,  which have been rescaled by the fundamental accruing time as $\tilde T \equiv T/\tau$, such that 
\begin{eqnarray}
\tilde T_{af}=M \left( 2^{K+1}-1 \right).
\end{eqnarray}

\subsection{Non-adaptive scheme}

Updating the control phase $\Phi$ without any dependence on the previous detection results requires that number of detections vary as a function of $k$.\cite{Higgins2009} From {Ref.\ \cite{Higgins2009}}, we choose a simple linear function for the number of detections,
\begin{eqnarray}
M(K,k)=M_K+F(K-k),
\end{eqnarray}
where $M_K$ is an initial number of detections at $k=K$, and $F$ is a positive integer. The number of detections increases as $k$ is decreased. Hence, the total time resource required to complete the protocol with the nonadaptive scheme is
\begin{eqnarray}
\tilde T_{na}=M \left( 2^{K+1}-1 \right) + F\left(2^{K+1}-2-K\right).
\end{eqnarray}
We determine the control phase $\Phi$ at the $m$-th {click} measurement as 
\begin{eqnarray}
\Phi_{m}=\Phi_{m-1}+\pi/2,
\label{predeterminedcontrol}
\end{eqnarray}
and use the same updating formula for the $b$'s as (\ref{updatingformulafortheprobabilityamplitude}).
Furthermore, the individual steps of this scheme are similar to those of the adaptive one, except for the following steps:
\begin{enumerate}
\item[1.]
All parameters are set: the fundamental accruing time $\tau$, the maximum number of multi-accumulations {$2^K$}, the initial number of detections $M_K$, the number of trials $\mathbf{S}$, and the dephasing time $T_2$.
\item[2.]
For the $\mathbf{s}$-th trial, the protocol starts using the initial values of $k=K$,  $M_K$, and $\Phi$.
\item[8.]
The phase estimate $\hat{\phi}$ is inferred and the control phase $\Phi$ is updated following (\ref{predeterminedcontrol}).
\item[9.]
Steps 3--8 are repeated for $M(K,k)=M_K+F(K-k)$ times.
\end{enumerate}

{In contrast to} the adaptive scheme, where one has to run the protocol from the maximum $k=K$ to the lowest $k=0$, the nonadaptive scheme can be operated without adhering to  the order. This flexibility could possibly simplify an experimental realization.

\begin{figure}[tp] 
\begin{center}
\setlength{\unitlength}{1cm}
\begin{picture}(6.5,6.5)
\put(0,0){\includegraphics[height=6.5\unitlength]{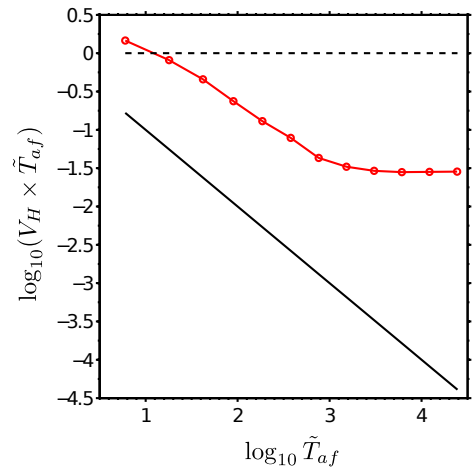}}
\end{picture}
\end{center}
\caption{(Color online) Logarithmic graphs of the total time resource $\tilde T_{af}$ versus the quantity $V_H \tilde T_{af}$ of the phase estimate. We plot the shot-noise scaling as the black dashed line and $\sim 1/\tilde T$ (Heisenberg-like) scaling as the black solid line. The numerical result of our protocol with adaptive feedback is plotted with a red dotted-solid line. For $\tilde T_{af} < \tilde T_2$, we achieve Heisenberg-like scaling of the estimate precision. We set the number of samples $\mathbf{S}=1000$, the number of single-shot detections $M=6$, $K=1\rightarrow20$, and the dephasing time $\tilde T_2= T_2/\tau=10^3$. For $\tilde T_{af} \geq 10^3$, the method fails and we can, at best, achieve shot-noise scaling through increasing the number of detections $M$ at a fixed $K$.}
\label{adaptive_ideal}
\end{figure}

\section{Simulation}

We numerically simulated the above protocol for the adaptive scheme, in the case of perfect SSM visibility, by setting the number of trials $\mathbf{S}=1000$ and the number of single-shot measurements $M=6$. The simulations were performed for values of $K$ from 1 to 20. Meanwhile, the ratio of dephasing time to $\tau$ is chosen to be $T_2/\tau=10^3$. A random phase guess is used as a prior estimate to initiate the protocol. The simulation result is presented in Fig. \ref{adaptive_ideal}, where we plot the simulated quantity $V_H \tilde T_{af}$ against $\tilde T_{af}$ on logarithmic scales. As can be seen from Fig. \ref{adaptive_ideal}, the precision of the simulated phase estimate {has Heisenberg-like scaling} for $\tilde T_{af} \leq 10^3$. If one proceeds with the estimation protocol for $\tilde T_{af} \geq 10^3$, the precision will be worse than the shot-noise scaling (not shown in the figure). Therefore, to maintain the shot-noise scaling for $\tilde T_{af} \geq 10^3$, one needs to stop the protocol at a certain value of $K$ that gives $T \approx T_2$, and subsequently increase $M$ only. It is {important to note that for these measurements,} no prior knowledge of the actual phase is needed.

\begin{figure}[tp] 
\begin{center}
\setlength{\unitlength}{1cm}
\begin{picture}(10,10)
\put(.3,0){\includegraphics[height=10\unitlength]{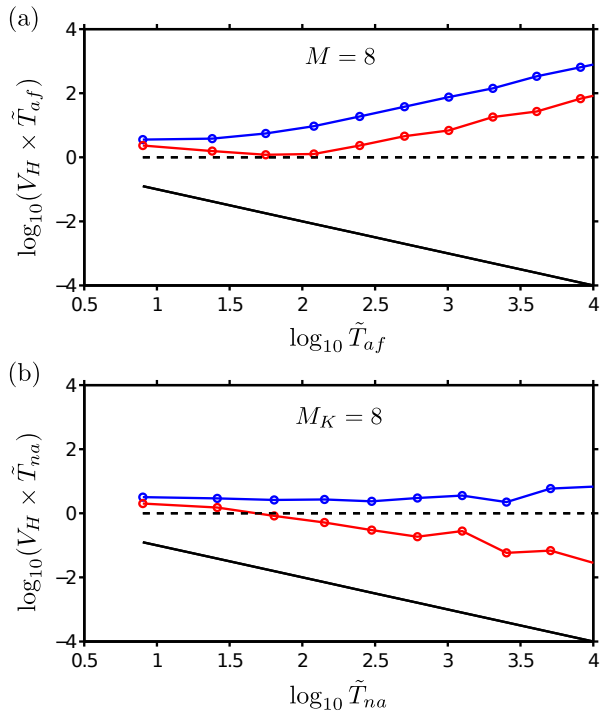}}
\end{picture}
\end{center}
\caption{(Color online) Logarithmic graphs of $\tilde T_{af}$($\tilde T_{na}$) versus $V_H \tilde T_{af}$ ($V_H \tilde T_{na}$) when the detection visibility is imperfect $f_d<1$, in the case of (a) the adaptive scheme, and (b) the nonadaptive scheme. The red (blue) dotted-solid lines set $f_d=95\%$ ($f_d=85\%$), and set $K=1\rightarrow20$, $\mathbf{S}=5000$, and $T_2/\tau=10^3$ for both simulations. We choose $F=2$ here for the nonadaptive method.}
\label{compare}
\end{figure}

\begin{figure}[tp] 
\begin{center}
\setlength{\unitlength}{1cm}
\begin{picture}(7,7)
\put(0.4,0){\includegraphics[height=7\unitlength]{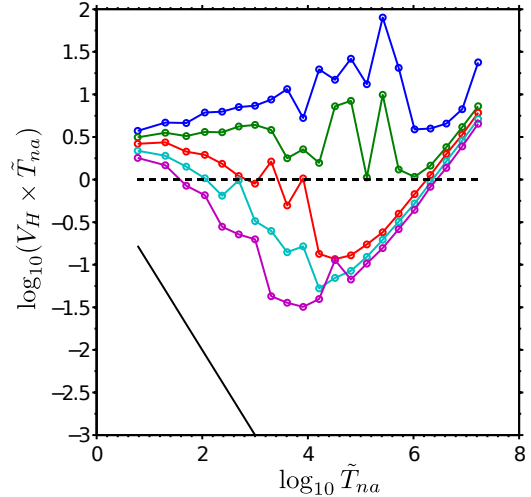}}
\end{picture}
\end{center}
\caption{(Color online) Logarithmic graphs of $\tilde T_{na}$ versus $V_H \tilde T_{na}$, for various detection visibility parameters: $f_d$= 82\% (blue), 86\% (green), 90\% (red), 94\% (cyan), and 98\% (magenta). The other parameters in the simulations are $T_2/\tau=10^3$, $M_K=6$, $K=1\rightarrow20$, $F=2$, and $\mathbf{S}=5000$.}
\label{nonadap3d}
\end{figure}

To consider a more realistic condition where a perfect single-shot measurement is unattainable, we quantify the imperfection both by a detection contrast and a detection visibility. The detection contrast of SSM is quantified by two parameters: a maximum and a minimum of the detection probability to find the spin state in an eigenstate of the detection basis. Here, we denote such maximum and minimum probability to get the $+1$ click by $f_{a}$ and $f_{i}$. Using these parameters, we straightforwardly generalize (\ref{DetectionProbFeedback}) into
\begin{eqnarray}
P(\pm 1 | \phi,\Phi,f_{a},f_{i}) = 
\frac{f_a+f_i}{2} 
\pm \frac{f_a-f_i}{2} {\cal Y},
\label{ContrastDetectionProbFeedback}
\end{eqnarray}
where ${\cal Y}=V(\tau_k) \cos(2^k\phi - \Phi)$,
giving the updating formula for the probability distribution coefficients as
\begin{eqnarray}
\tilde b_{j,m+1} &=& \left(f_a+f_i\right) b_{j,m} + \frac{u_{m+1}}{2} \left(f_a-f_i\right)V(\tau_k) \cr
&& \times \left(b_{j-2^k,m} e^{-i \Phi_m} + b_{j+2^k,m} e^{i \Phi_m} \right).
\label{updatingformulafortheprobabilityamplitudecontrast}
\end{eqnarray}
For the more symmetric case where $f_{a}+f_{i}=1$, we set $f_{a}-f_{i}=f_d$, 
and simplify (\ref{ContrastDetectionProbFeedback}) such that
\begin{eqnarray}
P(\pm 1 | \phi,\Phi,f_d)  = \frac{1}{2} \left( 1 \pm f_d V(\tau_k) \cos \left( 2^k\phi - \Phi \right)\right),
\label{GeneralisedDetectionProbFeedback}
\end{eqnarray}
giving the updating formula
\begin{eqnarray}
\tilde b_{j,m+1} &=& b_{j,m} + u_{m+1} f_d V(\tau_k) \nonumber \\
&& \times \left(  \frac{b_{j-2^k,m} e^{-i \Phi_m} + b_{j+2^k,m} e^{i \Phi_m}}{2} \right).
\label{updatingformulafortheprobabilityamplitudewithfd}
\end{eqnarray}

The detection visibility is included in our simulations to compare the robustness of both adaptive and nonadaptive schemes against the SSM imperfection. We {first choose parameters to optimize the nonadaptive protocol for a given detection visibility (i.e.\ the symmetric case where $f_{a}+f_{i}=1$). In the next subsection, we optimize the nonadaptive protocol to improve its robustness for imperfect detection contrast (the nonsymmetric case).

Using (\ref{GeneralisedDetectionProbFeedback}) and (\ref{updatingformulafortheprobabilityamplitudewithfd}), simulations were performed for a range of nonunit detection visibilities. As shown in Fig. \ref{compare}, it was found that the nonadaptive scheme is more robust against imperfect detection visibility. The results for the nonadaptive method with a range of visibilities are presented in Fig. \ref{nonadap3d}, indicating that a detection visibility of at least 90\% is required by the nonadaptive scheme to obtain Heisenberg-like scaling. While performing the simulations to obtain both Figs. \ref{compare} and \ref{nonadap3d}, we did not limit the value of $K$, in order to show that the variance has worse scaling than} the shot-noise limit for $\tilde T_{af,na}> \tilde T_2$. 

\subsection*{Optimized protocol with nonadaptive scheme}

\begin{figure}[tp] 
\begin{center}
\setlength{\unitlength}{1cm}
\begin{picture}(10.3,10.3)
\put(.15,0){\includegraphics[height=10.4\unitlength]{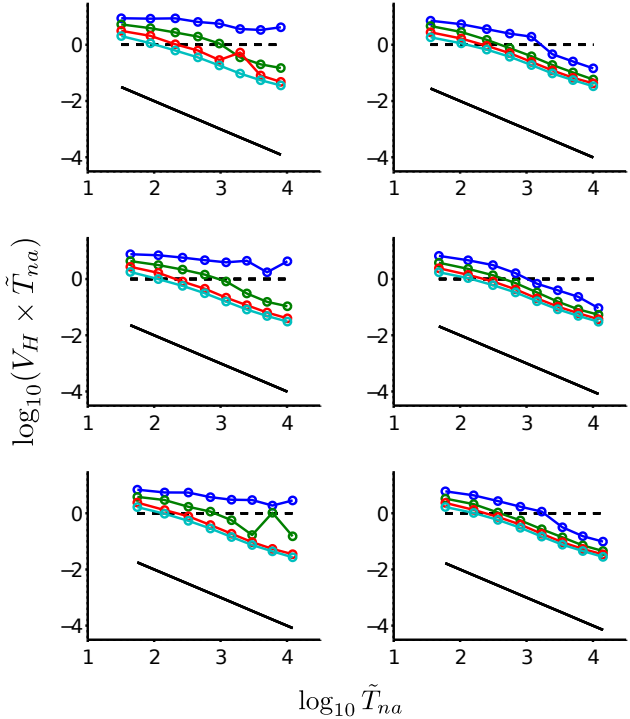}}
\end{picture}
\end{center}
\caption{(Color online) Logarithmic graphs of $\tilde T_{na}$ against $V_H \tilde T_{na}$, obtained from the optimized nonadaptive protocol for different detection contrasts and for some combinations of $\{M_K,F\} =$ $\{8,8\}$ (a), $\{12,8\}$ (b), $\{16,8\}$ (c), $\{8,12\}$ (d), $\{12,12\}$ (e) and $\{16,12\}$ (f). The detection contrast parameters used are $f_{i}=5\%$ and $f_{a}=55\%$(blue), $65\%$(green), $75\%$(red), and $85\%$(cyan). The black solid line represents {Heisenberg-like} scaling, while the shot-noise scaling is shown by the black dashed line. The simulations use 10000 samples, $T_2/\tau=10^3$, $K=1\rightarrow8$.
}
\label{optimalMF}
\end{figure}

The nonadaptive protocol can be optimized to work with nonideal detection contrast. We optimize the protocol by numerical simulations based on 
(\ref{ContrastDetectionProbFeedback}) and (\ref{updatingformulafortheprobabilityamplitudecontrast}) to find realizable combinations of the number of detections $M$, and the multiplier $F$ to maintain the sub-shot-noise scaling. This is useful to provide some optimized parameters for a possible experiment where the measurement efficiency is often not symmetric, as {quantified} by the detection contrast $\{f_a,f_i\}$.

We present the simulation results in Fig. \ref{optimalMF}, where the protocol is optimized for some values of $M_K=\{8,12,16\}$ and $F=\{8,12\}$, at a range of visibility parameters $f_a=\{0.55, 0.65, 0.75, 0.85\}$, where $f_i=0.05$ is fixed. 

From the figure, the overall results show the very surprising ability of the nonadaptive protocol to demonstrate near Heisenberg-like scaling ($\delta B\sim 1/T$), even at low detection contrasts ($f_a=0.55$). For low values of $(M_K, F)$, the method exhibits a scaling  which is worse than or comparable to the shot-noise limit. However, as these parameters are increased ($M_K, F=8\mapsto 12$), the method exhibits a significant improvement and moves toward $\sim 1/T$ scaling. In addition, this behavior occurs for a very large range of detection contrasts ($f_a>0.55$), and thus could be of great use to many other physical models besides the case of magnetometry here, where one does not have perfect single-shot readout.

\section{Conclusion}
We have shown that unambiguous magnetic field strength estimation {with sub-shot-noise scaling}
can be achieved by performing a time-varying sequence of measurements of the nuclear spin in a single N-V, using either adaptive or nonadaptive phase shifting of the system phase throughout the sequence.  By numerically simulating the protocol for both ideal and imperfect single-shot measurement conditions, we found the surprising result that the nonadaptive protocol, which is assumed to be far easier to realize experimentally, provides an estimate whose precision scales better than the adaptive protocol. Another remarkable feature of our proposed method is that by sampling the phase accumulation at varying times the protocol yields an unambiguous estimate of the magnetic field. This means that one can determine an absolute estimate of the local magnetic field strength, even though the field may cause many Larmor rotations of the spin. The sub-shot-noise scaling of the estimate precision is only bounded below by the Heisenberg scaling ($\delta B\sim 1/T$), and the unavoidable dephasing of the spin.

Recent experimental progress on coherent manipulations of the nuclear spin in the nitrogen-vacancy system in diamond allows one to perform SSM on the nuclear spin,\cite{Neumann2010b} and this may soon lead to experimental demonstrations of our protocol with subsequent benefits for large scale nanomagnetic imaging. For the same total imaging time, the precision of our proposed estimation technique scales better than that of the standard estimation method. Moreover, our proposal is applicable to any other atomic-like physical systems, such as an electron spin in silicon {(SSM in that system has recently been reported \cite{Morello2010})}, atomic vapors, or Bose-Einstein condensates.

\begin{acknowledgements}
We thank P. Hemmer, F. Jelezko, J. Beck, F. Rempp, T. Gaebel, C. Bradac, P. Kolenderski, and J. Sch\"{o}nfeldt for their useful discussions and comments. R. S. Said acknowledges Fields Institute for the travel fund during Thematic Program on Mathematics in Quantum Information 2009. This work was partially supported by the European Commission Project (Q-Essence).
\end{acknowledgements}

\end{document}